%% file: main.tex
\documentclass[manuscript,screen]{acmart}

\renewcommand\footnotetextcopyrightpermission[1]{}
\setcopyright{none}
\makeatletter
\fancypagestyle{plain}{
  \fancyhf{}            
  \fancyfoot[C]{\thepage}

}
\makeatother

\AtBeginDocument{%
  }

\usepackage{graphicx}
\usepackage{multirow}
\usepackage{multicol}
\usepackage{enumitem}
\usepackage{pifont}

\newcommand{\tool}[0]{PatchFuzz}%

\begin{document}

\title{\tool{}: Patch Fuzzing for JavaScript Engines}
\authorsaddresses{}
\author{Junjie Wang}
\email{junjie.wang@tju.edu.cn}
\affiliation{%
\institution{College of Intelligence and Computing, Tianjin University}
\country{China}
}
\author{Yuhan Ma}
\email{mayuhan@tju.edu.cn}
\affiliation{%
\institution{College of Intelligence and Computing, Tianjin University}
\country{China}
}
\author{Xiaofei Xie}
\email{xfxie@smu.edu.sg}
\affiliation{%
\institution{Singapore Management University}
\country{Singapore}
}
\author{Xiaoning Du}
\email{xiaoning.du@monash.edu}
\affiliation{%
\institution{Monash University}
\country{Australia}
}
\author{Xiangwei Zhang}
\email{xiangwei@tju.edu.cn}
\affiliation{%
\institution{College of Intelligence and Computing, Tianjin University}
\country{China}
}

\begin{abstract}
Patch fuzzing is a technique aimed at identifying vulnerabilities that arise from newly patched code.  
While researchers have made efforts to apply patch fuzzing to testing JavaScript engines with considerable success, these efforts have been limited to using ordinary test cases or publicly available vulnerability PoCs (Proof of Concepts) as seeds, and the sustainability of these approaches is hindered by the challenges associated with automating the PoC collection.
To address these limitations, we propose an end-to-end sustainable approach for JavaScript engine patch fuzzing, named \tool{}. 
It automates the collection of PoCs of a broader range of historical vulnerabilities and leverages both the PoCs and their corresponding patches to uncover new vulnerabilities more effectively.
\tool{} starts by recognizing git commits which intend to fix security bugs.
Subsequently, it extracts and processes PoCs from these commits to form the seeds for fuzzing, while utilizing code revisions to focus limited fuzzing resources on the more vulnerable code areas through selective instrumentation. 
The mutation strategy of \tool{} is also optimized to maximize the potential of the PoCs.
Experimental results demonstrate the effectiveness of \tool{}.  
Notably, 54 bugs across six popular JavaScript engines have been exposed and a total of \$62,500 bounties has been received.
\end{abstract}





\maketitle

\input{intro}

\input{motivation}
\input{preliminary}

\input{approach}
\input{evaluation}
\input{related}
\input{conclude}
\input{data}

\bibliographystyle{unsrtnat}
\bibliography{ref}

\end{document}

%% file: intro.tex
\section{Introduction}

JavaScript is broadly integrated into web applications, making it a primary threat of untrustworthy remote content to users' computing devices.
JavaScript engines in web browsers are responsible for parsing, executing, and optimizing JavaScript, and the fact that JavaScript language is dynamically and weakly typed makes it challenging for JavaScript engines to handle incoming input securely.
Once a JavaScript engine is compromised, remote attackers may gain control of the entire system without being detected.
Fuzzing, as a prevalent and effective method for testing complex software, is hired to automatically expose vulnerabilities in JavaScript engines.
To improve the effectiveness of JavaScript engine fuzzing, a significant effort has been devoted to diversifying the generated JavaScript samples~\cite{holler2012langfuzz, han2019codealchemist, aschermann2019nautilus, wangfuzzjit, gross2018fuzzil, park2020fuzzing, lee2020montage} and expanding the test coverage to include rarely executed code and execution paths \cite{bohme2017directed, luo2022selectfuzz, chen2018hawkeye}.

Meanwhile, another important line of research, known as \textit{patch fuzzing}, has emerged from testing newly patched code locations to uncover any remaining bugs that may not have been entirely fixed or may have been introduced by the patch \cite{wang2019superion, park2020fuzzing}.
This approach is prompted by the observation that a significant number of patches are incomplete or incorrect, and the presence of vulnerabilities can indicate complex code implementation, with additional vulnerabilities potentially in the code related to the patched area.
An early study~\cite{park2012empirical} found that 22\% to 33\% of resolved bugs necessitated more than one fix attempt.
Although this study did not specifically focus on JavaScript engines, similar scenarios arise due to the intricacy of JavaScript engines' code and bugs, and the practices of patch fuzzing for JavaScript engines have demonstrated considerable success \cite{gross2018fuzzil, wangfuzzjit, wang2019superion, park2020fuzzing}. 

At the core of patch fuzzing lies the imperative to generate high-quality test cases that can comprehensively evaluate the code patches. This necessitates test cases to not only access and engage with the patches effectively but also to undergo mutation to probe areas near the patched code.
Crafting trigger inputs for known vulnerabilities from scratch is an extremely challenging task. Nonetheless, if we manage to gather vulnerability PoCs (Proof of Concepts), this can provide a strong starting point for the fuzzer, enabling it to focus on regions where additional security bugs are more likely to be discovered.
This approach mirrors the strategies employed by many effective JavaScript engine fuzzing tools~\cite{han2019codealchemist,park2020fuzzing,lee2020montage}. 
To develop an efficient patch fuzzing tool, it is essential that the initial seeds function in harmony with the mutation and feedback mechanisms to maximize their potential.
In the following, we explore the current state of these three critical aspects and outline our strategies for addressing the challenges we encounter.

\textbf{Initial seed collection.} 
In the history of JavaScript engine fuzzing, researchers have attempted to collect PoCs from different resources, including the bug-revealing test cases generated by existing fuzzers~\cite{holler2012langfuzz}, the official ECMAScript conformance test suite~\cite{han2019codealchemist,lee2020montage}, e.g., Test262~\cite{test262}, the regression test suite of mainstream JavaScript engines~\cite{han2019codealchemist,park2020fuzzing}, and some manually collected CVE PoCs, e.g., Js-vuln-db~\cite{js-vuln-db}. 
Test262 and the regression test suite are still being actively maintained, but Js-vuln-db has not been updated since 2019 due to the substantial manual effort required for maintenance.
However, PoCs of historical vulnerabilities are more valuable in exposing security bugs.
Although security PoCs could be included in the regression test suite, the suite itself is a miscellaneous mixture of test cases for various testing purposes, including functionality and performance, with security PoCs only constituting a small portion due to their scarcity.
Using regression test cases as seeds could distract fuzzers from focusing on the most interesting areas.
Hence, none could offer complete and pure coverage of security PoCs of JavaScript engines.
\textbf{Can we find a sustainable way to gather security PoCs for JavaScript engines?}

Security bugs or vulnerabilities in open-source JavaScript engines are addressed through git commits.
Sometimes, their corresponding PoCs are included in the commits as test cases for future regression testing.
This key insight motivated us to establish a PoC collection from the vulnerability-fixing commits.
Identifying these security patch ones from the enormous commits is non-trivial since they have been made less distinguishable from others for some security reasons~\cite{wang2022graphspd}.
For example, to avoid the exposure of the relationship between a CVE and its corresponding patch, CVE IDs are often excluded from the commit messages.
CVE IDs indicate vulnerabilities with high security risk; if their patches are available, but turn out to be incomplete, it becomes more convenient for attackers to develop exploits and mount attacks.
Existing works revealed that these ``secretly'' fixed vulnerabilities can also be discerned, by analyzing commit messages and/or code revision with the help of machine learning techniques~\cite{zhou2021spi,wu2022enhancing,wang2022graphspd,nguyen2022vulcurator,nguyen2022hermes}.
These approaches have been tested as effective on a group of open-source projects.
However, in our evaluation of the open-source JavaScript engines, their performance experienced a significant drop, mainly due to the inherent generalization problem of machine learning models.
Therefore, we manually analyzed a small group of JavaScript engine commits and heuristically developed a set of recognition rules.
Our evaluation showed that this approach achieved an average precision and recall of 0.89 and 0.97, respectively, enabling us to better identify vulnerability-fixing commits and extract any available PoCs.

\textbf{Revision-based feedback.}
In patch fuzzing, we assume that additional vulnerabilities may be present in the patches or in code that shares functionalities similar to the original problematic code. Our primary objective is to generate test cases that predominantly focus on these code patches, rather than other uninterested code snippets. This leads us to a question: \textbf{Can we leverage the code revision information in security patches (e.g., vulnerability-fixing commits) to better guide the patch fuzzing?} While this challenge may seem akin to existing directed fuzzing approaches as AFLGo~\cite{bohme2017directed}, there are fundamental differences. Directed fuzzing mainly targets ``never-reached'' lines of code, whereas in our scenario, we already possess PoCs that cover the desired code lines. Additionally, due to the lack of available PoCs, conventional directed fuzzing relies on intensive static analysis to incrementally select test cases that gradually approach the target lines, achieving new coverage. The directed fuzzing approaches may not scale efficiently for large projects like JavaScript engines. In light of these considerations and the collection of initial PoCs, we propose a simple but more scalable approach: a selective instrumentation strategy. This strategy prioritizes test cases that achieve new coverage in the target lines of the code, aligning more closely with the specific requirements of patch fuzzing in large-scale projects.

\textbf{Mutation intensity.}
For the sake of targeted fuzzing, it is also important to
not smash the original semantics of PoCs but ensure adequate mutations to exercise more code. 
Wild mutations can destroy the original PoCs and make not much difference from using random seed.
However, if the mutations are too conservative, it limits the fuzzers from covering more executions.
As far as we know, DIE~\cite{park2020fuzzing} is the first work acknowledging the importance of PoC integrity and intends to maintain their integrity during mutation. 
However, its strategies have been too conservative where no alterations are allowed for the control flow structures. For example, some control structures, e.g., loops, are critical to triggering bugs.
As a consequence, it is unable to detect certain bug cases, as exemplified in our motivating example, i.e., CVE-2018-8137, in Figure~\ref{fig:CVE-2018-0777}, which introduced a new bug triggered by adding an additional layer of \texttt{for} loop to the original PoC of CVE-2018-0777.
\textbf{Hence, appropriately controlling the mutations on PoCs is critical.} In this paper, we propose an appropriate mutation strategy for patch fuzzing of JavaScript engines.

In summary, to effectively identify vulnerabilities in JavaScript engines that are introduced or overlooked by security patches, this paper presents an end-to-end sustainable security patch fuzzing approach for open-source JavaScript engines, named \tool{}.
It automates the entire fuzzing lifecycle from identifying security patches from the git commits to generating the PoCs triggering bugs.
In particular, it overcomes the key problems mentioned earlier: 
1) the extraction of executable PoCs from security patches by identifying security patches for JavaScript engines, 2) effective and efficient guidance for fuzzing toward the exploration of the patched code and its surroundings, and 3) an appropriate mutation strategy to balance the integrity and explorability of PoCs.

For recognizing security patches, we derived a set of heuristic rules by manually analyzing a small set of git commits.
Next, two key assets, PoCs and code revision locations, are extracted from these security patches.
The PoCs are not directly executable for JavaScript engines and we summarize a comprehensive list of rules to facilitate the automatic transformation.
Afterward, they are leveraged as seeds for fuzzing in order to generate penetrating test cases for evaluating the effectiveness of software patches.
In light of optimizing the allocation of resources, we further selectively instrument the targeted JavaScript engines in accordance with the call graph of security patches.
For the first time, PoCs are utilized in conjunction with code revisions in security patches for fuzzing. 
Finally, in order to maximize the efficacy of PoCs, we employ a mutation technique that combines two levels of intensity and properly adapts the newly introduced variables to the current context.

In comparison with the baseline approaches, \tool{} demonstrates significant improvement in precision and recall for identifying security patches and detects an increased number of bugs. 
We also evaluate the importance of each step in \tool{} through an ablation study.
In particular, \tool{} successfully identifies 54 bugs within six mainstream JavaScript engines, which results in the assignment of 25 CVEs and an impressive payout of \$62,500 for bug bounties.

In summary, this work offers the following contributions:
\begin{itemize}[noitemsep,topsep=0pt]
\item An innovative way to identify and process PoCs from security patches via detecting security patches for JavaScript engines. 
\item The largest security PoCs dataset of high quality for JavaScript engines.
\item An end-to-end sustainable patch fuzzing of JavaScript engines, leveraging selective instrumentation as a means to improve its effectiveness.
\item A comprehensive evaluation on the efficacy of \tool{} in identifying software vulnerabilities, with 54 bugs uncovered in six widely adopted JavaScript engines.
\end{itemize}

%% file: motivation.tex
\section{Motivating example}

To better motivate our research, we present a real-world example of how incomplete security patches can lead to the discovery of more vulnerabilities in JavaScript engines. 

\begin{figure}[htbp]
\centering
\includegraphics[width=0.7\linewidth]{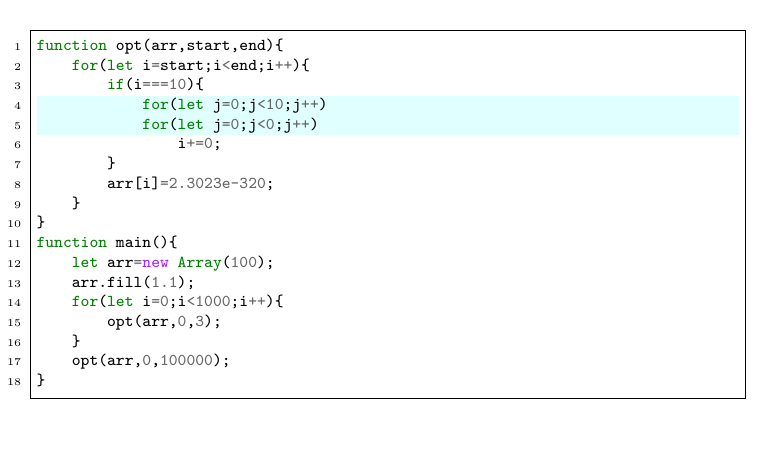}
\caption{
The PoC of CVE-2018-0777 (without line 4 and 5), 
the PoC of CVE-2018-8137 (with line 4 added to the PoC of CVE-2018-0777),
and a bypass to CVE-2018-8137 (with line 5 added to the PoC of CVE-2018-0777).
}
\label{fig:CVE-2018-0777}
\end{figure}

CVE-2018-0777~\cite{CVE-2018-0777} is a buffer overflow vulnerability that was found in ChakraCore, the JavaScript engine that was previously used by the Edge browser. 
This vulnerability was discovered by Lokihardt and was caused by an implementation bug in the JIT (Just-in-Time) compiler module. 
Lokihardt released a PoC in his bug report, which is displayed in Figure~\ref{fig:CVE-2018-0777}. 
The JIT compiler is responsible for the dynamic compilation and optimization of JavaScript code during runtime. 
Due to the complexity of the optimization logic that the erroneous code was intended to implement, CVE-2018-0777 is just the tip of the iceberg. 
The patch to fix it is far from complete, and two additional vulnerabilities were quickly discovered.
The PoCs for the two follow-up vulnerabilities can be found in Figure~\ref{fig:CVE-2018-0777}. 
The differences between the three PoCs are minor as highlighted, which demonstrates the feasibility of detecting new vulnerabilities with the help of PoCs and small perturbations.

Next, to unfold the inspirations obtained from this case, we have to delve into the root cause of each vulnerability and analyze how its patch can be bypassed easily.
The bugs in the case are mainly related to the 
implementation errors in deciding if \textit{Bound Check Elimination} can be applied for optimization, especially when it is entangled with two other optimization techniques, \textit{Constant Folding} and \textit{Dead Code Elimination}.
We will briefly explain each technique along with the PoCs.

The compilation and optimization techniques are to speed up the execution of the JavaScript input.
They are usually triggered when some code is executed many times.
In Figure~\ref{fig:CVE-2018-0777}, the \texttt{main} function hires a \texttt{for} loop (14 to 16) to trigger the compilation and optimization of the function \texttt{opt}.
However, an important bound check in \texttt{opt} is incorrectly removed. 
Later, when \texttt{opt} is invoked again at line 17, the incorrect assembly code is executed and a buffer overflow happens.

In \texttt{opt}, there is an array indexing operation inside the \texttt{for} loop at line 8.
The JavaScript engine will check if the index is within the valid bound of an array before each indexing.
However, when the index is verified to be always valid, the check can be eliminated to speed up the execution.
Depending on whether the index is an induction variable, the verification operates in different ways.
Induction variables are those that get updated with a fixed offset during each loop iteration.
The value of an induction variable in the $N$-th iteration is represented as $s+c*N$, where $s$ is its initial value, and $c$ is the offset that will be added to it after each iteration.
For an array of length $len$, if $s+c*N\leq len-1$ is always true for any $s$, $N$, and $len$, the upper bound checking can be safely omitted.
In this example, \texttt{i} would be an induction variable if it were not updated by lines 3 to 7.
These lines add more complexity to \texttt{i}'s dataflow, making it hard to decide if \texttt{i} remains an induction variable.
However, the addition assignment at line 6 is optimized to be an assignment \texttt{i = 10} with constant folding, as whenever this statement is executed \texttt{i} must be 10.
However, this optimization sets $N$ to be \texttt{-0x80000000} but fails to update \texttt{i} to be non-inductive. 
As a result, the bound check elimination is still verified against the formula $s+c*N\leq len-1$.
While given $c$ is 1, $N$ is \texttt{-0x80000000}, $s$ is an integer, and $len$ are non-negative, $s+c*N\leq len-1$ will always hold, since the maximum value for an integer is \texttt{0x7fffffff}, while the minimum value for $len$ is $0$, and \texttt{0x7fffffff-0x80000000 = -1 $\leq$ 0-1} holds.

The patch for CVE-2018-0777 fixes the problem by updating the status of \texttt{i} to be non-inductive, but it does not cover all the cases during constant folding optimization, such as when a loop is involved during the folding. 
As a consequence, a bypass was discovered by a researcher from Tencent Zhanlulab~\cite{lzhenhuan} by adding only one statement to the original PoC, as shown in Figure~\ref{fig:CVE-2018-0777}.
The bug is identified as CVE-2018-8137~\cite{CVE-2018-8137}.
Later, even after the second patch, the problem fails to be completely mitigated.
Another bypass is developed based on the PoC of CVE-2018-8137, by making the updating to \texttt{i} at line 6 to be dead code.
Here it triggers the optimization of dead code elimination, which distracts the JIT module from executing the patches, and the bound check is incorrectly removed again~\cite{lzhenhuan}.

\noindent \textbf{Remarks.} There are several important takeaways from this example. 
Firstly, vulnerabilities are a good indicator of complex code while fixing a vulnerability in complex code is challenging and may require several attempts.
Second, PoCs of historical vulnerabilities are an extraordinary starting point to discover vulnerabilities overlooked or introduced by their patches.
Third, code that is functionally related to a vulnerability is also likely to be vulnerable, and more fuzzing power shall be scheduled to exercise it.
Last, when mutating the historical PoCs, it is essential not to destroy their penetrating capability in order to better explore the sibling paths or deeper paths of the previously buggy path.
In this example, it is important to retain the \texttt{for} loops and the \texttt{if} statement at line 3 because otherwise the JIT module would not be triggered for the bound check elimination and constant folding.

%% file: preliminary.tex
\section{Preliminary study}\label{sec:preliminary}

One of the primary obstacles in patch fuzzing is the collection of security PoCs, which guide the fuzzing tool directly to the specific lines of code affected by a vulnerability. 
This task is generally difficult. 
Nevertheless, based on our observations of security maintenance in open-source JavaScript engines, we have found that security PoCs can be retrieved from git commits used to patch security bugs. 
Consequently, the task becomes to first identify security patches among numerous miscellaneous git commits.

Previous studies have made attempts to identify security patches by analyzing the git commit messages~\cite{zhou2021spi,wu2022enhancing,wang2022graphspd,nguyen2022vulcurator,nguyen2022hermes, wang2021patchdb,tan2021locating}, code revisions~\cite{zhou2021spi,wu2022enhancing,wang2022graphspd,nguyen2022vulcurator,nguyen2022hermes,wang2021patchdb,tan2021locating}, and issue reports~\cite{nguyen2022vulcurator,nguyen2022hermes}, using various neural networks.
However, none of these models have been evaluated for detecting security patches of JavaScript engines. 
These learning-based approaches rely on labeled datasets for training and evaluation. 
To the best of our knowledge, there are no labeled datasets available for security patch commits of JavaScript engines. 
Therefore, in this preliminary study, we first establish a small-scale dataset of JavaScript engine commits and manually label whether they are security patches or not. 
Afterwards, we assess the performance of SOTA security patch identifiers on the curated datasets.

\noindent\textbf{Dataset preparation.}
In this study, we selected six widely used open-source JavaScript engines, namely JavaScriptCore (JSC)~\cite{jsc}, V8~\cite{V8}, SpiderMonkey (SM)~\cite{sm}, ChakraCore (CH)~\cite{ch}, JerryScript (Jerry)~\cite{jerryscript}, and QuickJS (QJS)~\cite{qjs} to construct the datasets.
These engines also serve as the subjects of our experimental evaluation, and you can find more detailed information about them in Table~\ref{tab:targets}.

Previous works~\cite{nguyen2022vulcurator, zhou2021spi, wang2022graphspd, wu2022enhancing, nguyen2022hermes, wang2021patchdb,tan2021locating} have primarily relied on datasets labeled according to the CVE patch information maintained in databases like NVD. However, this approach may overlook numerous security patches that are not indexed in these databases due to unavailable information or cases where the bugs they fix have not been assigned a CVE. 
To address this limitation, researchers have attempted manual labeling of commits~\cite{zhou2021spi, sabetta2018practical}, but this endeavor was only feasible for a limited number of open-source projects due to the significant efforts involved.

JavaScript engines typically maintain their bug records using their own Bugzilla or issue lists and only a small fraction of bugs were assigned CVE IDs. 
Upon examining the CVEs associated with JavaScript engines over the past three years, we discovered that few were accompanied by patch information. 
Consequently, we decided to manually label a small set of commits for each engine.
To minimize inaccuracies stemming from manual labeling methods, the first two authors, possessing more than five years of experience in C/C++ development and security analysis, independently labeled the commits. 
Later, they had a discussion to reconcile any discrepancies in their results. 
For each JavaScript engine, we randomly selected historical commits and requested the two authors to label them until we obtained 50 security patch commits and 50 negative ones. Ultimately, we amassed 100 labeled samples for each engine.

\noindent\textbf{Performance evaluation on baselines.}
Among the existing security patch recognition tools~\cite{zhou2021spi,wu2022enhancing,wang2022graphspd,nguyen2022vulcurator,nguyen2022hermes},
we selected VulCurator~\cite{nguyen2022vulcurator} and GraphSPD~\cite{wang2022graphspd} as our baselines because they are the SOTA approaches and have been open-sourced.
Specifically, VulCurator is a deep learning-based technique designed to assess git commit messages, code revisions, and issue reports. 
Its evaluation on the SAP dataset~\cite{sabetta2018practical} and a self-curated TensorFlow~\cite{abadi2016tensorflow} dataset yielded F1 scores of 0.79 and 0.89, respectively. 
The SAP dataset consists of 1,132 manually-labeled security patches and 5,995 non-security patches, sourced from 220 Java and Python projects. 
The TensorFlow dataset comprises 290 security patches labeled with the NVD database and 1,535 non-security patches obtained from the TensorFlow repository. 
Negative samples in both datasets were randomly sampled from non-security patch commits. 
VulCurator utilized 80\% of each dataset for training and 20\% for testing.
GraphSPD is a graph neural network based security patch detection system, which represents patches as graphs with semantics and utilizes a patch-tailored graph model for detection.

We applied VulCurator and GraphSPD to our JavaScript engine datasets and measured the precision, recall, and F1 score for each engine. 
According to the results in Table~\ref{tab:vulcurator}, a surprisingly poor average F1 score of 0.09 of VulCurator is observed, with an average precision of 0.78 and an average recall of only 0.05. 
This outcome can likely be attributed to two factors: 1) VulCurator's limited generalizability, as its training data is relatively small in scale and originates from different domains, and 2) the bias of the TensorFlow dataset toward vulnerabilities indexed in NVD, which may result in inadequate coverage of other security patches.
When applying GraphSPD to our JavaScript engine datasets, we found that it failed to generate CPGs (Code Property Graphs) for the majority of git commits, specifically 395 out of 600 patches. 
For the remaining 205 git commits for which it successfully generated CPGs, the average precision was 0.50, and the average recall was 0.03.

\begin{table}[htbp]
\centering
\scriptsize
\caption{The performance of VulCurator and GraphSPD on JavaScript engine git commits classification.}
\label{tab:vulcurator}
\begin{tabular}{c|c|cccccc|c}
\toprule
Baseline & Metric & JSC & V8 & SM & CH & Jerry & QJS & Average \\
\cmidrule{1-9}
& Precision & 1.00 & 1.00 & 0.92 & 1.00 & 0.00 & 0.76 & 0.78 \\
VulCurator & Recall & 0.03 & 0.01 & 0.12 & 0.09 & 0.00 & 0.05 & 0.05 \\
& F1 & 0.06 & 0.02 & 0.21 & 0.17 & 0.00 & 0.08 & 0.09\\
\cmidrule{1-9}
& Precision & 1.00  & 0.00 & 1.00& 0.50 & 0.00 & 0.50 & 0.50 \\
GraphSPD & Recall & 0.06 & 0.00  & 0.04 & 0.04 & 0.00 & 0.04 & 0.03 \\
& F1 & 0.11 & 0.00 & 0.07 & 0.04  & 0.00 & 0.02 & 0.04 \\
\bottomrule
\end{tabular}
\end{table}

%% file: approach.tex
\section{Approach}

\begin{figure*}[htbp]
\centering
\includegraphics[width=\linewidth]{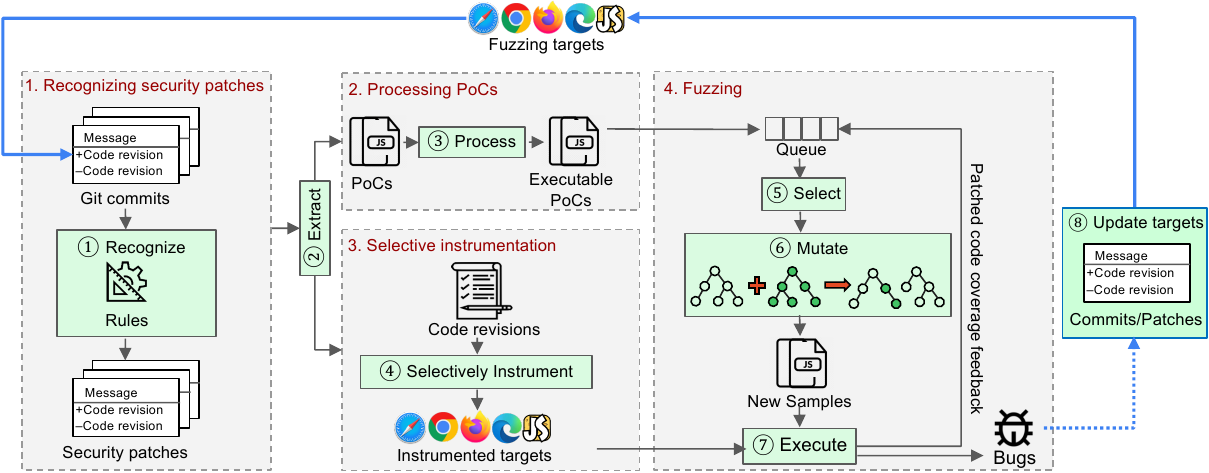}
\caption{The general workflow of \tool{}.}
\label{fig:workflow}
\end{figure*}

In this section, we present our end-to-end sustainable patch fuzzing approach for JavaScript engines, called \tool{}, along with its workflow illustrated in Figure~\ref{fig:workflow}.
When provided with a specific version of a JavaScript engine as the fuzzing target, \tool{} automatically identifies commits that address security vulnerabilities (Step \ding{172}).
From these commits, \tool{} further extracts executable PoCs as well as the code changes (Step \ding{173}).
What sets \tool{} apart is its unique utilization of both the extracted PoCs and the code changes to design effective patch fuzzing tools (Step \ding{174}).
Leveraging the power of both, \tool{} directly focuses on vulnerable areas and deliberately exercises the surrounding code with specialized code coverage guidance (Step \ding{175}).
Furthermore, the mutation strategy employed by \tool{} is optimized to preserve the essential properties of the PoCs, ensuring their ability to penetrate vulnerabilities effectively (Steps \ding{176}-\ding{178}).
Whenever the fuzzing target is updated with new commits, \tool{} seamlessly adapts to the changes, moving on to test the updated version while incorporating the insights provided by the new commits (Step \ding{179}).
In the following sections, we present the details of all the primary design aspects of \tool{}.

\subsection{Recognizing security patches}

Security patches of JavaScript engines exhibit distinct patterns in their commit messages and code changes.
Hence, we developed a rule-based security patch recognizer in this research. 
The approach is simple and efficient and demonstrates a high level of accuracy in the evaluation.
Specifically, we determine whether a git commit is a security patch based on the following criteria.

\noindent\textbf{Associated bug report.}
In many cases, the maintainer of the JavaScript engine will provide the link or ID of the issue report that is fixed by the current git commit in its message or in the annotation of its code changes.
In the corresponding issue report, there is usually a specific attribute that indicates whether the issue is a security concern.
Therefore, we determine whether the git commit is a security patch by referring to the attribute in the issue report that indicates if it is a security issue.

\noindent\textbf{Keywords.}
In the message or the annotations of code changes of a git commit, there are often certain keywords, such as \texttt{crash}, \texttt{buffer overflow}, \texttt{integer overflow}, and so on, that indicate whether the git commit is a security patch.
To this end, we maintain a list of security-related keywords and use the presence of these keywords as an indicator of a security patch.

\noindent\textbf{Conditional statement.}
In security patches, large sections of new code are typically not introduced as they are in feature patches.
In many cases, security patches will add or modify a conditional statement to enhance security checks.
For example, they might check the boundary values of certain variables, whether pointer variables are null, if operations have caused integer overflow, if memory allocation is successful, or if certain variables have been initialized.
Therefore, we consider the addition or modification of conditional statements in code changes as a characteristic of security patches. 
However, since regular feature patches also frequently use conditional statements, we calculate the ratio of the number of added or modified conditional statements to the total number of modified lines of code as a distinguishing feature.

\noindent\textbf{Attached test case.}
To verify whether a security patch has successfully fixed a vulnerability, some git commits include test cases. 
However, some commits that fix high-risk vulnerabilities might not include test cases. 
Additionally, feature patches also include test cases to verify whether functional errors have been resolved.
However, there are differences between the contents of security test cases and functional test cases. 
Functional test cases tend to focus on whether the computation results are correct or if the performance meets certain benchmarks, while security test cases are more focused on determining whether the test cases lead to crashes or other security issues.

\subsection{Extracting executable PoCs}

In this section, we present our approach for extracting and processing PoCs from security patches.
Our motivation for the processing step stems from the observation that over a quarter of the extracted PoCs are unable to be executed successfully by JavaScript engines.
Next, we present the details of the study and how it inspires us to design a rule-based PoC processing tool. 

\noindent\textbf{Extracting PoCs.}
PoCs typically exist as JavaScript files, which sets them apart from the source code of JavaScript engines written in C/C++ language.
Moreover, to facilitate regression testing, the PoCs, along with other test cases, are organized and stored in specific folders.
Hence, we gather information on the addition and modification of JavaScript files within security patches and analyze the corresponding folders used to house them.
It is important to note that not all JavaScript files in these folders qualify as PoCs, as many are intended for normal testing purposes such as functional and stress testing. 
Including a large number of non-PoC files as seeds may distract the fuzzing process, leading it away from focusing on the vulnerable areas of the JavaScript engines.

\noindent\textbf{A usability study of extracted PoCs.}
The seeds of JavaScript engines are expected to be free of grammar errors, including syntax errors that occur when the JavaScript code violates the language's syntax rules, as well as reference errors that arise when a variable referenced has never been declared or is not in scope.
However, the PoCs extracted from security patches often undergo modifications to facilitate their compatibility with the test runner for automated regression testing. 
As a result, these modified PoCs may reveal grammar errors when directly fed into JavaScript engines.
To determine the presence of grammar errors in these PoCs, we leverage the exit code generated during their execution. 
The exit code, also referred to as the return code, is a value that a program or command returns to the operating system upon completion of its execution. 
It is usually represented as a non-negative integer, where a value of \texttt{0} indicates successful execution, while non-zero values indicate various types of errors.
Among the exit codes, certain values are indicative of grammar errors. 
For instance, \texttt{3} is used to indicate grammar errors by JavaScriptCore and SpiderMonkey, while \texttt{1} is used by V8, ChakraCore, JerryScript, and QuickJS.

\begin{figure}[htbp]
  \centering
  
  \begin{minipage}[t]{0.45\textwidth}
    \centering
    \includegraphics[width=\linewidth]{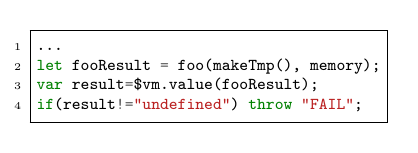}
    \caption{PoC of CVE-2018-4416 before processing}
    \label{fig:CVE-2018-4416-2}
  \end{minipage}\hfill
  \begin{minipage}[t]{0.45\textwidth}
    \centering
    \includegraphics[width=\linewidth]{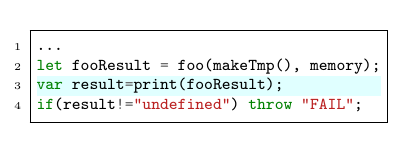}
    \caption{PoC of CVE-2018-4416 after processing}
    \label{fig:CVE-2018-4416-3}
  \end{minipage}
  
\end{figure}

We conduct a dry run to collect the exit codes for all the gathered PoCs and subsequently categorize the errors based on their exit code and error messages.
The grammar error rates of the collected PoCs for each JavaScript engine are presented in Table~\ref{tab:poc_rules}. 
They range from 10.9\% to 64.3\%, with an average of 27.5\%.
We also present some of the recurring errors specific to each JavaScript engine, along with the proportion of each error among all the errors.
It is noteworthy that reference errors constitute the largest proportion across all engines. 
These errors are typically associated with the special identifiers required by test runners.
In Figure~\ref{fig:CVE-2018-4416-2}, we illustrate an example PoC that triggers a reference error due to the absence of the object definition for \texttt{\$vm} during execution. 
The \texttt{\$vm} object is designed to evaluate the state of the JavaScriptCore virtual machine during runtime but is not supported in the default JavaScriptCore engine. 
Next, we introduce a rule-based approach to eliminate each type of error.

\begin{table}[htbp]
\centering
\scriptsize
\caption{The grammar error in extracted PoCs and processing rules.}
\label{tab:poc_rules}
\begin{tabular}{cclrc}
\toprule
Target & Overall error rate & Error & Ratio & \tool{}\\
\cmidrule{1-5}
\multirow{3}{*}{JSC} & \multirow{3}{*}{11.1\%} & ReferenceError: \$vm & 56.72\% & Rewrite \\
& & ReferenceError: load & 29.61\% & Remove \\
& & SyntaxError: export & 2.28\% & Remove\\
\cmidrule{1-5}
\multirow{2}{*}{V8} & \multirow{2}{*}{64.3\%} & ReferenceError: assert* & 61.22\% & Rewrite \\
& & ReferenceError: load & 16.54\% & Remove\\
\cmidrule{1-5}
\multirow{4}{*}{SM} & \multirow{4}{*}{27.2\%} & ReferenceError: libdir & 42.42\% & Remove\\
& & ReferenceError: Wasm-related & 7.16\% & Rewrite \\
& & ReferenceError: appendToActual & 5.14\% & Rewrite\\
& & ReferenceError: assert* & 2.07\% & Rewrite\\
\cmidrule{1-5}
\multirow{3}{*}{CH} & \multirow{3}{*}{22.9\%} & ReferenceError: testRunner & 39.90\% & Rewrite\\
& & ReferenceError: telemetryLog & 7.88\% & Rewrite\\
& & ReferenceError: assert & 3.94\% & Rewrite\\
\cmidrule{1-5}
Jerry & 10.9\% & ReferenceError: export & 4.10\% & Remove\\
\cmidrule{1-5}
QJS & 28.6\% & ReferenceError:\_\_loadScript & 92.86\% & Rewrite\\
\cmidrule{1-5}
Average & 27.5\% & - & - & - \\
\bottomrule
\end{tabular}
\end{table}

\noindent\textbf{Rule-based PoCs processing.}
Depending on the level of entanglement between the statements causing the grammar errors and other code lines through control flow or data flow, we employ different strategies to address them. 
If the error-causing code is relatively independent, we simply remove the specific code lines or tokens responsible for the error. 
However, if the code lines are interconnected with the rest of the code, we aim to find harmless replacements for the erroneous tokens.
For instance, in the case of the reference error depicted in Figure~\ref{fig:CVE-2018-4416-2}, we replace the unrecognized \texttt{\$vm.value} with a \texttt{print} method, as illustrated in Figure~\ref{fig:CVE-2018-4416-3}.
Although this substitution alters the original semantics, it allows the subsequent code to be executed successfully.
The fixing strategies for the errors in Table~\ref{tab:poc_rules} are also listed in the last column.
It is important to note that our approach does not resolve all the errors, as some of them require substantial efforts to fix. 
Nonetheless, it effectively reduces the overall error rates from 27.5\% to 8.4\% (refer to Section~\ref{subsec:recognizer} for details), making it the most effective solution to our best knowledge.

\subsection{Selective instrumentation}

In this section, we introduce the study carried out on code revisions identified from security patches and how the findings are leveraged to design the selective instrumentation for improving the efficiency of \tool{}.

\noindent\textbf{Hot code study.}
Despite the widespread belief that vulnerabilities are present in a small portion of a large code base~\cite{Zimmermann2010,du2019leopard}, there has been a lack of quantitative studies conducted on JavaScript engines, mainly due to the challenges associated with identifying security patches. 
In this study, leveraging the capabilities of \tool{}, we present a quantitative analysis of the portion of source files that are patched against security bugs and how frequently they are patched.
The frequently patched code is also known as \textit{hot} code.

Table~\ref{tab:changed_files} displays the total number of source files in each JavaScript engine and the number of files modified at least once in security patches.
On average, approximately 35.8\% of the source files have undergone patches to address security vulnerabilities, with each file being patched an average of 33.7 times.
These findings align with previous observations made by other researchers~\cite{Zimmermann2010,du2019leopard}, where vulnerabilities tend to be concentrated and recurring in a small number of files.
Notably, JavaScriptCore, V8, and SpiderMonkey exhibit lower ratios, with percentages below 8.9\%, while ChakraCore, JerryScript, and QuickJS demonstrate considerably higher ratios above 36.7\%.
Further investigation into the potential factors influencing these ratios uncovered that ChakraCore has implemented a file-level modularization approach, resulting in larger files and increased maintenance complexity. 
On the other hand, JerryScript and QuickJS remain relatively immature due to insufficient development and maintenance, with vulnerabilities spotted in many files.

\begin{table}[htbp]
\footnotesize
\centering
\caption{The number of changed files in security patches.}
\label{tab:changed_files}
\begin{tabular}{l|rrrrrr|r}
\toprule
Target & JSC & V8 & SM & CH & Jerry & QJS & Average\\
\cmidrule{1-8}
\#Source files & 32,865 & 27,525 & 37,117 & 1,778 & 608 & 30 & -\\
\#Changed files & 2,939 & 1,707 & 1,798 & 1,565 & 426 & 11 & -\\
Ratio(\%) & 8.9 & 6.2 & 4.8 & 88.0 & 70.1 & 36.7 & 35.8\\
Average patched times & 33.9 & 47.3 & 78.2 & 4.2 & 4.9 & 33.7 & 33.7\\
\bottomrule
\end{tabular}
\end{table}

\noindent\textbf{Selective instrumentation.}
In existing fuzzing methods for JavaScript engines, code coverage guidance is typically achieved by instrumenting all source code files. 
However, due to the large and complex nature of JavaScript engines and the slow execution speed of test samples, it becomes crucial to prioritize limited time and resources toward testing specific code locations that are more likely to contain vulnerabilities. 
Based on the insights gained from the above study, we aim to focus on fuzzing the hot code through selective instrumentation. 
We instrument all the functions that are directly changed in the security patches or functions that have a calling relationship with these functions.
To compute the calling relationship between functions, we construct the call graph of the target JavaScript engines by applying Doxygen~\cite{Doxygen}.
Then we instrument all functions that call or are called by patched functions.


Our selective instrumentation approach helps increase the efficiency of security patch testing in two ways. 
Firstly, selective instrumentation can reduce the number of instrumentation points, thereby reducing the computational workload for code coverage analysis of each sample and improving the running efficiency of fuzzing. 
Secondly, only new samples that trigger new code coverage in security patches will be added to the fuzzing queue, while samples that exercise less important functional modules, such as checking for syntax errors in samples, will not be saved. 
This can help the fuzzing process to focus on exploring areas where security vulnerabilities have occurred in the past.

\subsection{Mutation}

To more effectively control the mutation intensity of fuzzing, ensuring that the generated test cases do not compromise the original exploitability of the PoCs while still being capable of discovering new vulnerabilities, \tool{} proposes a combination of two different mutation intensities.
One is a low-intensity mutation strategy, where each change to the sample is relatively small and only involves expression-level mutations. 
The other is a high-intensity mutation method, where each change to the sample is more substantial and involves statement-level mutations.
Specifically, in the low-intensity mutation strategy, we randomly select an expression from the test sample each time and either delete it, replace it with an expression from the same sample or another sample, or insert an expression from the same sample or another sample at a random position to generate a new sample.
This mutation strategy helps us discover vulnerabilities such as the one shown in Figure~\ref{fig:CVE-2018-0777} that bypass patches for CVE-2018-8137.
In the high-intensity mutation method, we randomly select a statement from the test sample each time and either delete it, replace it with a statement from the same sample or another sample, or insert a statement from the same sample or another sample at a random position between existing statements to generate a new sample.
This mutation strategy helps us discover vulnerabilities similar to CVE-2018-8137, as shown in Figure~\ref{fig:CVE-2018-0777}.
In our experiments, we use the low-intensity mutation method with an 80\% probability and the high-intensity mutation method with a 20\% probability.

During the sample mutation process, it is necessary to adapt the variables in the newly introduced expressions or statements to the new context. 
Otherwise, the newly generated samples are likely to trigger syntax errors.
Therefore, our strategy is to replace the variable names in the newly introduced expressions or statements with the variable names from the sample to be mutated. 
This ensures that the newly generated samples do not encounter undefined variable errors.
When replacing variable names, it is also necessary to consider the variable types. 
Replacing with a variable of the wrong type can lead to type errors.
As JavaScript is a weakly-typed and dynamically-typed language, it is difficult to determine the types of variables through static methods. 
To obtain the types of variables, we choose to insert probe code and dynamically execute it to determine the runtime types of variables in the sample.
Specifically, we first parse the sample to obtain a list of all variables that appear in it. 
Then, we insert a piece of probe code between each statement. 
This probe code first determines whether each variable from the complete list of variables is accessible at the current position by checking if it is \texttt{undefined}. 
For accessible variables, it then determines their specific primitive or object types by reading their \texttt{constructor.name}.

%% file: evaluation.tex
\section{Evaluation}

\noindent In order to evaluate the effectiveness of \tool{} and the importance of each component, we design experiments to answer the following research questions and present a case study on a zero-day vulnerability discovered by \tool{}.
\begin{itemize}[noitemsep,topsep=0pt]
\item \textbf{RQ1}: How is the performance of \tool{} in identifying security patches and extracting executable PoCs? 
\item \textbf{RQ2}: What is the effectiveness of the extracted PoCs in enhancing the performance of various fuzzing tools?
\item \textbf{RQ3}: How effective is \tool{} in exposing new vulnerabilities? What about comparing with SOTA techniques?
\item \textbf{RQ4}: Does every component in \tool{} contribute to its effectiveness?
\end{itemize}

\noindent\textbf{Implementation and settings.}
We implemented the security patch identification, PoC extraction, and seed mutation in Python 3.
ANTLR 4~\cite{antlr} is hired to facilitate the parsing and mutation of JavaScript.
Additionally, the binary level coverage-guided fuzzing is powered by AFL (American Fuzzy Lop)~\cite{afl}, and we use AFL-cov~\cite{afl-cov} for the calculation of source level code coverage.
For selective instrumentation, we created a shell script to instruct AFL to instrument the target source code functions.
All experiments were conducted on a workstation equipped with an Intel i9-9900K processor that boasts 8 cores. 
The workstation is also equipped with 32GB of RAM and the Ubuntu 18.04 operating system.

\noindent\textbf{Testing targets.} 
To evaluate our approach, we selected six mainstream JavaScript engines, namely JavaScriptCore, V8, SpiderMonkey, ChakraCore, JerryScript, and QuickJS.
Specifically, JavaScriptCore, V8, and SpiderMonkey are the JavaScript engines for Safari, Chrome, and Firefox browsers, respectively. 
ChakraCore was the JavaScript engine for the Edge browser, but Edge replaced it with V8 in March 2021.
Hence, our experiments on ChakraCore only cover the period before it was abandoned.
JerryScript is a renowned JavaScript engine specifically designed and implemented for IoT devices.
QuickJS is a small and embeddable Javascript engine.

We summarize basic information about the six JavaScript engines in Table~\ref{tab:targets}, including the number of lines of code and the fuzzing throughput (i.e., the number of samples executed per second).
V8 has the largest amount of code while JerryScript has the smallest.
On average, there are 365,014 lines of code. 
The execution speed of different engines also varies a lot, with the fastest one, QuickJS, executing 315.1 samples per second, while the slowest one, SpiderMonkey, executes just 1.9 samples per second. 

\begin{table}[htbp]
\footnotesize
\centering
\caption{The general information of targets, commits collected, and the number of security patches recognized.}
\label{tab:targets}
\begin{tabular}{c|rrccrrr}
\toprule
Target & \#Lines & Throughput(s) & Start date & End date & \#Commits & \#Security patchs & Ratio(\%)\\
\cmidrule{1-8}
JSC & 279,449 & 135.6 & 2001/08/24 & 2024/06/30 & 262,194 & 21,720 & 8.3\\
V8 & 975,453 & 33.7 & 2008/07/03 & 2024/06/30 & 80,730 & 10,927 & 13.5\\
SM & 540,580 & 1.9 & 1998/03/28 & 2024/06/30 & 835,209 & 75,656 & 9.1\\
CH & 272,210 & 46.5 & 2016/01/05 & 2024/06/30 & 13,033 & 2,434 & 18.7\\
Jerry & 33,905 & 299.6 & 2014/07/01 & 2024/06/30 & 4,539 & 718 & 15.8\\
QJS & 88,491 & 315.1 & 2020/09/07 & 2024/06/30 & 170 & 28 & 16.5\\
\cmidrule{1-8}
Average & 365,014 & 138.7 & - & - & 199,312 & 18,580 & 13.7\\
\bottomrule
\end{tabular}
\end{table}

\noindent\textbf{Baselines.}
We compare \tool{} with the SOTA general-purpose fuzzing tool, Fuzzilli~\cite{gross2018fuzzil}, FuzzJIT~\cite{wangfuzzjit}, and two other premium fuzzers utilizing historical test cases, Superion~\cite{wang2019superion} and DIE~\cite{park2020fuzzing}. 
Fuzzilli and FuzzJIT employ a hybrid approach of generational and mutational fuzzing without requiring initial seeds. 
All four tools instrument the entire code base to collect feedback on code coverage. 
Superion and DIE perform mutations at the AST level, while Fuzzilli and FuzzJIT operate at a finer-grained level of an intermediate language.
We were unable to apply Fuzzilli to ChakraCore and DIE and FuzzJIT to JerryScript and QuickJS due to a lack of fuzzing interface.
Also, DIE failed to generate any new samples for ChakraCore after the fuzzing campaign successfully started.
This may be due to the incompatibility between the interface provided by DIE three years ago and the current changes in ChakraCore's code. 

\subsection{\textbf{RQ1:} Security patch recognition and executable PoCs extraction}\label{subsec:recognizer}

\noindent\textbf{Recognizing security patches.}
We reuse the 600 labeled samples prepared in the Preliminary study (Section~\ref{sec:preliminary}) to evaluate our rule-based security patch recognizer and report the precision, recall, F1 score, and accuracy.
We present the results of each engine in Table~\ref{tab:security_patch} and report the average.
It can be seen that our approach substantially outperforms the baselines, VulCurator and GraphSPD (see Table~\ref{tab:vulcurator}).
\tool{} achieves an accuracy of 0.92 and a high F1 score of 0.93, with a precision of 0.89 and a recall of 0.97.

\begin{table}[htbp]
\centering
\footnotesize
\setlength{\tabcolsep}{6pt}
\caption{The performance of security patches recognizer.}
\label{tab:security_patch}
\begin{tabular}{c|cccc|c}
\toprule
\multirow{1.5}[3]{*}{Target} & \multicolumn{4}{c|}{Controlled evaluation} & Wild evaluation\\
\cmidrule{2-6}
& Precision & Recall & F1 & Accuracy & Accuracy\\
\cmidrule{1-6}
JSC     & 0.91 & 0.98 & 0.94 & 0.94 & 0.95 \\
V8      & 0.98 & 1.00 & 0.99 & 0.99 & 0.98 \\
SM      & 0.94 & 0.98 & 0.96 & 0.96 & 0.97 \\
CH      & 0.80 & 0.94 & 0.86 & 0.85 & 0.88 \\
Jerry   & 0.84 & 0.94 & 0.89 & 0.88 & 0.88 \\
QJS     & 0.87 & 0.98 & 0.94 & 0.94 & 0.92 \\
\cmidrule{1-6}
Average & 0.89 & 0.97 & 0.93 & 0.92 & 0.93 \\
\bottomrule
\end{tabular}
\end{table}

We present the statistics of the historical commits for each JavaScript engine in Table~\ref{tab:targets} and employ \tool{} to identify security patches from the corpus.
The details of the recognized security patches, including their numbers and ratios, are also provided in Table~\ref{tab:targets}.
Remarkably, an average of 13.7\% of the total collected commits are classified as security patches. 

We further evaluate the performance of \tool{} by manually validating a sampled set of the classified commits.
For each JavaScript engine, we randomly sampled 50 commits from both the recognized security patches and non-security patches. 
The first two authors scrutinized these commits, verifying their correct classification. 
The overall accuracy of this validation is reported in Table~\ref{tab:security_patch}.
It can be seen that the performance of our approach remained consistent with the results obtained from the controlled evaluation, which reinforces the reliability of the approach in accurately identifying and classifying git commits.

\noindent\textbf{Extracting executable PoCs.}
We apply \tool{} to process the test cases obtained from the identified security patches, transforming them into executable PoCs.  
In Table~\ref{tab:Grammar_error}, we present the number of extracted PoCs and their respective grammar error rates before and after processing.
For comparison purposes, we replicated DIE's algorithm for test case processing and assessed the grammar error rate subsequent to its execution. 
Notably, DIE does not offer a solution for processing JerryScript and QuickJS, so that particular measurements are omitted.
Observing the results, \tool{} effectively mitigates a significant portion of grammar errors in the PoCs, reducing the average error rate from 27.5\% to 8.4\%. 
Conversely, DIE's processing exhibits lower efficacy, and in certain cases, it even amplifies the grammar error rates (e.g., for JavaScriptCore and SpiderMonkey).
We delved into the cause of this disparity and discovered that it stems from DIE's outdated assumption regarding the compatibility of certain JavaScript engine features, such as \texttt{BigInt}.

\vspace{5pt}
\noindent
\fbox{
\begin{minipage}{0.97\textwidth}
\textbf{Answer to RQ1:}
The experimental results demonstrate that \tool{} exhibits superior performance in identifying security patches when compared with the baselines (i.e., VulCurator and GraphSPD) in terms of precision, recall, F1 score, and accuracy.
At the same time, \tool{} makes more PoCs executable by eliminating their grammar errors.
\end{minipage}
}

\begin{table}[htbp]
\footnotesize
\centering
\caption{The grammar error rate of PoCs before and after processing of \tool{} and DIE.}
\label{tab:Grammar_error}
\begin{tabular}{l|rrrrrr|r}
\toprule
Target & JSC & V8 & SM & CH & Jerry & QJS & Total\\
\cmidrule{1-8}
\#PoCs & 3,952 & 2,895 & 7,450 & 2,434 & 718 & 28 & 17,477\\
Rrror rate & 11.1\% & 64.3\% & 27.2\% & 22.9\% & 10.9\% & 28.6\% & 27.5\%\\
\cmidrule{1-8}
\tool{} & 6.8\% & 2.9\% & 13.6\% & 12.9\% & 10.4\% & 3.6\% & 8.4\%\\
DIE & 12.3\% & 15.5\%  & 39.2\% & 16.3\% & - & - & - \\
\bottomrule
\end{tabular}
\end{table}

\subsection{\textbf{RQ2:} Effectiveness of extracted PoCs}

To assess the usefulness and effectiveness of our extracted PoCs, we conducted a comparative analysis against three existing baseline seed collections for JavaScript engines: Js-vuln-db~\cite{js-vuln-db}, Test262~\cite{test262}, and DIE's corpus~\cite{park2020fuzzing}. 
We feed these four distinct seed corpora to three fuzzers requiring initial seeds: Superion~\cite{wang2019superion}, DIE~\cite{park2020fuzzing}, and \tool{}. 
Each JavaScript engine target was subjected to a week-long fuzzing run for each seed corpus, allowing us to calculate the average number of bugs detected and the average line coverage. 
This process was repeated 3 times for each configuration, and was conducted on the latest versions of the target JavaScript engines.

The results are shown in Table~\ref{tab:seeds_coverage}. 
The first column lists the various seed corpora and their respective sizes. 
The `Target' column indicates the JavaScript engines tested, while `Count' shows the number of seeds used for each target. 
It is noteworthy that the seeds from DIE's corpus and our PoCs are collected to each engine, leading to a varied number of seeds used per engine. 
Column `Initial Coverage' reflects the initial coverage attained by the seeds before fuzzing, and columns `\#Bug' and `Coverage' display the average bug detection and line coverage, respectively.

\begin{table}[htbp]
\footnotesize
\centering
\caption{The bug-revealing ability and line coverage comparison with baseline seed collections.}
\label{tab:seeds_coverage}
\begin{tabular}{c|cc|c|cccccc}
\toprule
\multirow{1.5}[2]{*}{Seeds} & \multirow{1.5}[2]{*}{Target} & \multirow{1.5}[2]{*}{Count} & \multirow{1.5}[2]{*}{Initial Coverage} & \multicolumn{2}{c}{Superion} & \multicolumn{2}{c}{DIE} & \multicolumn{2}{c}{\tool{}}\\
\cmidrule{5-10}
& & & & \#Bug & Coverage & \#Bug & Coverage & \#Bug & Coverage \\
\midrule
& JSC & 182 & 0.23 & 1.3 & 0.28 & 0.3 & 0.29 & 1.3 & 0.31 \\
& V8 & 182 & 0.11 & 0.0 & 0.14 & 0.0 & 0.14 & 0.0 & 0.19\\
& SM & 182 & 0.19 & 0.0 & 0.20 & 0.0 & 0.19 & 0.0 & 0.25\\
Js-vuln-db & CH & 182 & 0.30 & 0.3 & 0.34 & - & - & 1.0 & 0.38\\
182 & Jerry & 182 & 0.51 & 0.7 & 0.75 & - & - & 0.7 & 0.74\\
& QJS & 182 & 0.28 & 0.7 & 0.33 & - & - & 0.6 & 0.35\\
\cmidrule{2-10}
& Average & 182 & 0.27 & 0.5 & 0.34 & \textbf{0.1} & 0.21 & 0.6 & 0.37\\
\midrule
& JSC & 28,580 & 0.24 & 0.0 & 0.29 & 0.0 & 0.29 & 0.0 & 0.31\\
& V8 & 28,580 & 0.10 & 0.0 & 0.13 & 0.0 & 0.12 & 0.0 & 0.18\\
& SM & 28,580 & 0.23  & 0.0 & 0.24 & 0.0 & 0.23 & 0.0 & 0.30\\
Test262 & CH & 28,580 & 0.33  & 0.0 & 0.37 & - & - & 0.0 & 0.42\\
28,580 & Jerry & 28,580 & 0.76  & 0.0 & 0.76 & - & - & 0.0 & 0.80\\
& QJS & 28,580 & 0.32 & 0.0 & 0.37 & - & - & 0.0 & 0.39\\
\cmidrule{2-10}
& Average & 28,580 & 0.33 & 0.0 & 0.36 & 0.0 & 0.22 & 0.0 & 0.40 \\
\midrule
& JSC & 3,673 & 0.34 & 0.3 & 0.42 & 0.3 & 0.42 & 0.6 & 0.45\\
& V8 & 3,641 & 0.14 & 0.0 & 0.18 & 0.0 & 0.19 & 0.0 & 0.24\\
 & SM & 5,167 & 0.31 & 0.0 & 0.33 & 0.0 & 0.31 & 0.0 & 0.41\\
DIE corpus & CH & 1,766 & 0.39 & 0.0 & 0.44 & - & - & 0.3 & 0.43\\
14,247 & Jerry & - & - & - & - & - & - & - & -\\
& QJS & - & - & - & - & - & - & - & -\\
\cmidrule{2-10}
& Average & 3,561 & 0.30 & 0.1 & 0.34 & \textbf{0.1} & 0.30 & 0.2 & 0.38 \\
\midrule
& JSC & 3,952 & 0.38 & 1.6 & 0.47 & 0.3 & 0.47 & 2.3 & 0.64\\
& V8 & 2,895 & 0.16 & 0.0 & 0.20 & 0.0 & 0.19 & 0.3 & 0.33\\
& SM & 7,450 & 0.38 & 0.0 & 0.39 & 0.0 & 0.38 & 0.0 & 0.63\\
Our PoCs & CH & 2,434 & 0.42 & 0.7 & 0.47 & - & - & 1.0  & 0.6\\
17,477 & Jerry & 718 & 0.78 & 1.0 & 0.81 & - & - & 1.0 & 0.86\\
& QJS & 28 & 0.46 & 0.9 & 0.48 & - & - & 0.8 & 0.66\\
\cmidrule{2-10}
& Average & 1,912 & \textbf{0.43} & \textbf{0.7} & \textbf{0.47} & \textbf{0.1} & \textbf{0.35} & \textbf{0.9} & \textbf{0.62} \\
\bottomrule
\end{tabular}
\end{table}

Overall, the results lead to several key findings: 
1) Our extracted PoCs achieve higher initial coverage compared to other corpora, despite being smaller in size than Test262 corpus. 
2) When examining the performance of different seed corpora across fuzzers, our PoCs consistently outperform others in bug detection and coverage. 
For instance, with our PoCs, Superion records (0.7, 0.47) for bug detection and coverage, surpassing the next best results of 0.5 bugs with Js-vuln-db and 0.36 line coverage with Test262. 
3) A tool-wise comparison using identical initial seeds reveals \tool{}’s superiority in bug detection and coverage. 
For example, using the DIE corpus, \tool{} achieves (0.2, 0.38), outstripping Superion and DIE, which achieve (0.1, 0.34) and (0.1, 0.30), respectively.
These results underscore the effectiveness of our extracted PoCs and \tool{} in enhancing bug detection and coverage compared to conventional baseline seed collections.

\vspace{5pt}
\noindent
\fbox{
\begin{minipage}{0.97\textwidth}
\textbf{Answer to RQ2:}
The extracted PoCs significantly boost the performance of existing fuzzers, enabling them to uncover more bugs and attain higher line coverage than existing seed collections.
\tool{} in particular shows exceptional results, outperforming Superion and DIE with the same seed corpus.
\end{minipage}
}

\subsection{\textbf{RQ3:} New bugs found and comparison with baselines}

\noindent\textbf{Experiment setting.}
We further evaluate the bug-finding capability of \tool{} and report the bugs detected.
It is important to note that the design of \tool{} has undergone extensive optimization and improvement over a prolonged period.
Here we report all the vulnerabilities detected by \tool{} ever since its very initial stage.
For the purpose of evaluation and comparison, a group of JavaScript engine versions (i.e., 48 versions) is curated as the testing targets, including those on which beta-version \tool{} used to expose vulnerabilities and the first versions of each calendar month ever since the start of 2019.
We re-execute the full-gear \tool{} on these versions to verify its ability to reproduce all vulnerabilities within a 7-day timeframe.
To establish a comparison, we executed baseline fuzzers on the same versions for an equivalent duration.
In addition to reporting newly discovered bugs, we measured the code coverage achieved by \tool{} and other baselines. 
Specifically, we assessed the line coverage across the entire codebase, as well as the coverage of frequently patched code files. 
Note that, in this experiment, \tool{}, Superion~\cite{wang2019superion} and DIE~\cite{park2020fuzzing} used the same seed corpus, i.e., our collected PoCs.

To minimize the errors caused by the randomness of fuzzing, we ran each experiment 10 times. 
Cumulatively, we conducted comprehensive and reliable evaluation, totaling 2,400 weeks of fuzzing, i.e., 5 tools $\times$ 10 random $\times$ 48 versions $\times$ 1 week.
We calculated the average value, and tested the statistical significance of \tool{} achieving better performance (i.e., higher coverage) than the baselines/variants with Vargha Delaney $\hat{A}_{12}$ and Mann Whitney U test (\textit{U}).
The Mann Whitney U test assesses if one list of observations is statistically greater than another, while the $\hat{A}_{12}$ statistic quantifies the size of the difference (effect size). 
A significant performance gap exists when the $p$-value of \textit{U} ($p_U$) falls below 0.05. 
If $\hat{A}_{12}$ is $\geq$ 0.71, it indicates that \tool{} outperforms significantly with a substantial effect size.

\noindent\textbf{New bugs found.}
Table~\ref{tab:bugs} provides an overview of the vulnerabilities detected by \tool{} and the baselines. 
The table utilizes a `\checkmark' to denote successful identification, an `x' to indicate a failure, and a `-' when fuzzing was not feasible for a specific target.
In total, \tool{} uncovered 54 vulnerabilities across the six JavaScript engines, with 21, 7, 3, 3, 13, and 7 identified in JavaScriptCore, V8, SpiderMonkey, ChakraCore, JerryScript, and QuickJS, respectively. 
For the baselines, Superion, DIE, and Fuzzilli detected 9, 2 and 5 vulnerabilities across all targets, respectively, all of which were also discovered by \tool{}. 
FuzzJIT fails to find any new crash bugs.
FuzzJIT does not rely on historical PoCs and is particularly designed for fuzzing the JIT modules, an extremely challenging task.
Out of the 54 vulnerabilities identified by \tool{}, 25 have been assigned CVE IDs, resulting in a total bounty of \$62,500 being awarded.
It is worth noting that the relatively smaller number of vulnerabilities discovered in ChakraCore should not be misconstrued as inherent superiority in security. 
Fuzzing for ChakraCore was halted after it was deprecated by Edge in March 2021, and thus, further vulnerabilities may have remained undetected.

\input{bugs2}

\noindent\textbf{Coverage.}
The coverage measurements (presented in Table~\ref{tab:coverage}) reveal the superior performance of \tool{} compared to all the baselines across each fuzzing target in terms of overall code coverage and hot code coverage.
The `Improvement' rows in the table indicate the coverage enhancement achieved by \tool{} in comparison to the baselines. 
Statistical significance tests, specifically the \textit{U} and $\hat{A}_{12}$ tests, confirm that all the improvements and their effect sizes are statistically significant.
When compared to the second-best performing fuzzers, namely Superion and Fuzzilli, \tool{} achieves approximately a 30\% higher coverage on both metrics.

\input{coverage2}

\vspace{5pt}
\noindent
\fbox{
\begin{minipage}{0.97\textwidth}
\textbf{Answer to RQ3:}
\tool{} has successfully uncovered 54 vulnerabilities, whereby 25 of them have been assigned a CVE number, and a total bounty of \$62,500 was awarded. 
Moreover, under experimental conditions, \tool{} has demonstrated superior effectiveness compared to baselines in discovering new vulnerabilities and achieving higher code coverage.
\end{minipage}
}

\subsection{\textbf{RQ4:} Ablation study}
\noindent\textbf{Experiment setting.}
To validate the effectiveness of \tool{}'s design, we conducted an ablation study that focused on three key aspects: seed collection, selective instrumentation, and mutation.
We created three distinct configurations of \tool{} by selectively disabling one design aspect at a time while keeping the remaining features intact.
In our previous analysis (RQ2), we have demonstrated the effectiveness of our seed collection. 
In this ablation study, we conducted additional evaluations using an alternative seed source, i.e., the DIE corpus (referred to as `-Seed').
For the selective instrumentation, we conducted additional evaluation using the full instrumentation of target JavaScript engines (referred to as `-Selective').
While for the mutation, we use the mutation strategy of DIE as a comparison (referred to as `-Mutation').
Following the same experimental setup as in RQ3, we assessed the ability of these \tool{} variants to reproduce the 54 vulnerabilities within a 7-day timeframe.

\noindent\textbf{New bugs found and coverage.}
The findings from the evaluation are presented in the corresponding sections of Table~\ref{tab:bugs} and Table~\ref{tab:coverage}. 
Notably, significant decreases in the number of reproduced vulnerabilities and code coverage were observed when each feature of \tool{} was disabled, underscoring the importance of each component.
Specifically, when the seeds were replaced, \tool{} only managed to reproduce 2 vulnerabilities. 
Similarly, disabling the selective instrumentation feature resulted in the discovery of only 10 vulnerabilities. 
Replacing the mutation strategy with DIE's strategy resulted in the discovery of only 6 vulnerabilities.
Furthermore, disabling the seeds had a severe impact on code coverage, resulting in a 44\% decrease in overall code coverage and a 50\% decrease in hot code coverage. 
Disabling other features also caused a loss of approximately 9\% to 11\% in overall code coverage and 3\% to 15\% in hot code coverage.

\vspace{5pt}
\noindent
\fbox{
\begin{minipage}{0.97\textwidth}
\textbf{Answer to RQ4:}
The removal of any design element within \tool{} results in a decrease in the number of detected bugs and code coverage, highlighting the crucial role played by each design aspect in enhancing \tool's performance. 
Notably, the seed collection has the most significant impact on bug detection, while the selective instrumentation has a comparatively lighter impact.
\end{minipage}
}

\subsection{Case study}

\begin{figure}[htbp]
\centering
\includegraphics[width=0.7\linewidth]{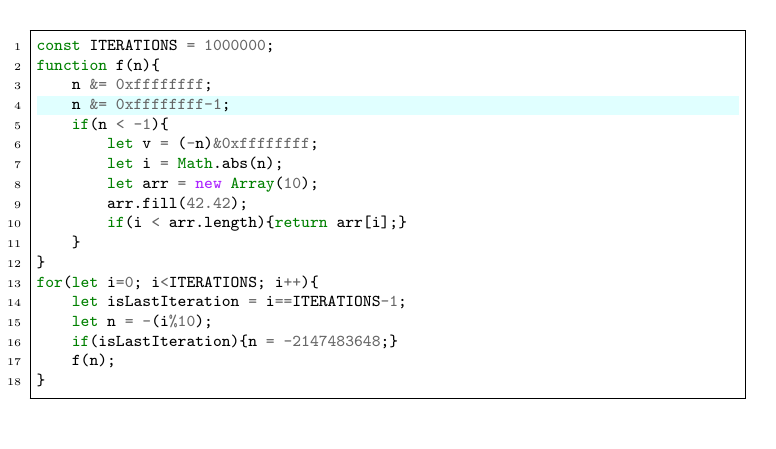}
\caption{The PoC of CVE-2020-9802 (without line 4) and the PoC of CVE-2020-9983 (without line 3).}
\label{fig:CVE-2020-9802}
\end{figure}

Validating the relationship between the vulnerabilities discovered by \tool{} and historical PoCs can be a challenging task.
In this section, we present a case study focusing on a buffer overflow vulnerability detected by \tool{} in JavaScriptCore. 
This case study highlights the significant role of \tool{} in assessing the security of hot code. 
The vulnerability causes the same threat as CVE-2020-9802~\cite{CVE-2020-9802}, which has been previously proven to be exploitable~\cite{CVE-2020-9802-exp}. 
We provide the PoC of CVE-2020-9802 and the new PoC derived from the original one through mutation in Figure~\ref{fig:CVE-2020-9802}.
The mutation involved a simple addition of the expression \texttt{-1} to line 4. 
This conservative mutation strategy employed by \tool{} aims to preserve the vulnerability-penetrating characteristics of the seed samples.
In the following, we delve into the root cause of CVE-2020-9802 and uncover how a similar but more hidden vulnerability is detected with the help of its PoC.  

CVE-2020-9802 was a result of an implementation error in the \textit{Common Subexpression Elimination} optimization within JavaScriptCore. 
This error led to an integer overflow, followed by a buffer overflow.
In Figure~\ref{fig:CVE-2020-9802}, the \texttt{for} loop (lines 13 to 18) triggers the optimization of the \texttt{f} function, which is invoked inside the loop at line 17. 
Two important optimizations are applied to \texttt{f}: \textit{Common Subexpression Elimination} at line 7 and \textit{Bound Check Elimination} at line 10. 
These optimizations are completed during the loop iterations, and the optimized assembly code is used when \texttt{f} is subsequently invoked.

Within the \texttt{if} branch at line 5, the variable \texttt{n} is guaranteed negative. 
As a result, the \texttt{Math.abs} operation on \texttt{n} at line 7 is equivalent to a negation operation.  
Negations should be checked for integer overflow to ensure the correctness of the result. 
Nevertheless, since \texttt{n} is forced to be a 32-bit integer at line 4 and the negation operation at line 6 is followed up with a \textit{BitAnd} operation to take the lower 32 bits of the result, an integer overflow check becomes unnecessary for this case.
The optimizer erroneously considers the calculation of negation at line 6 and the \texttt{Math.abs} operation at line 7 as common subexpressions and consequently eliminates the latter by simply copying the result from the former. 
As a result, the negation operation with an overflow check is replaced with one without check.
When \texttt{n} becomes \texttt{-2147473648} (i.e., \texttt{-0x80000000}), whose negation should be \texttt{0x80000000}, we need a 64-bit integer to store the negation result because it exceeds the upper bound of a 32-bit signed integer. 
However, since the integer overflow check is mistakenly removed, the variable \texttt{i} is assigned the value of \texttt{v}, which is \texttt{-2147473648}.
Later, at line 10, \texttt{i} is used for array indexing. 
Since \texttt{n} is checked to be negative and its absolute value \texttt{i} is assumed to always be positive, the bounds check for line 10 is removed as an optimization. 
Consequently, when \texttt{i} is incorrectly calculated as negative, an out-of-bounds access occurs.

The patch for CVE-2020-9802 addressed the issue by disallowing the replacement of a checked negation with an unchecked negation. 
However, in this case, when determining whether an integer overflow could occur in the negation operation at line 7, the range of \texttt{n} calculated after the \textit{BitAnd} operation at line 4 is incorrect. 
It mistakenly assumes that it will never cause an overflow.
It turns out that the calculation is consistently incorrect when the operand of the \textit{BitAnd} operation is negative, except for \texttt{0xffffffff}. 
Due to the incorrect elimination of the overflow check, the buffer overflow occurs once again.

One of the most concerning risks presented by such ``follow-up'' vulnerabilities is their potential to reuse the exploits, if accessible, from the original vulnerabilities to carry out attacks. 
The presence of historical bugs also provides a shortcut for identifying these new vulnerabilities. 
Consequently, we emphasize the need for increased community focus on patch fuzzing and raising the technical obstacles for adversaries attempting to exploit vulnerabilities linked to past bugs.

%% file: bugs2.tex
\begin{table*}[htbp]
\footnotesize
\centering
\setlength{\tabcolsep}{4.8pt}
\caption{Unique bugs detected during 7-days by \tool{}, baselines, and \tool{} variants.}
\label{tab:bugs}
\begin{tabular}{lcccr|ccccc|ccc}
\toprule
\multirow{2}{*}{\#} & \multirow{2}{*}{ID} & \multirow{2}{*}{Target} & \multirow{2}{*}{Status} & \multirow{2}{*}{Bounty} & \multicolumn{5}{c|}{Comparison with baselines} & \multicolumn{3}{c}{Ablation study}\\
\cmidrule{6-13}
 &                                  &       &           &           & \tool{}   & Superion  & DIE       & Fuzzilli  & FuzzJIT   & -Seed     & -Selective& -Mutation\\
\cmidrule{1-13}
1 & CVE-2019-7292                   & JSC   & fixed     & 0         & \checkmark& x         & x         & x         & x         & x         & x         & x         \\
2 & CVE-2019-8743                   & JSC   & fixed     & 0         & \checkmark& \checkmark& x         & x         & x         & x         & x         & x         \\
3 & CVE-2020-9983                   & JSC   & fixed     & 0         & \checkmark& x         & x         & x         & x         & x         & x         & x         \\
4 & CVE-2021-1817                   & JSC   & fixed     & 0         & \checkmark& x         & x          & x         & x         & x         & x         & x         \\
5 & CVE-2023-35074                  & JSC   & fixed     & 0         & \checkmark& x         & x         & x         & x         & x         & x         & x         \\
6 & bug-200388                      & JSC   & fixed     & 0         & \checkmark& \checkmark& x         & \checkmark& x         & \checkmark& \checkmark& \checkmark\\
7 & bug-215841                     & JSC   & fixed     & 0         & \checkmark& \checkmark& \checkmark & x         & x         & x         & x         & x         \\
8 & bug-216214                      & JSC   & fixed     & 0         & \checkmark& x         & x         & x         & x         & x         & \checkmark& \checkmark\\
9 & bug-225094                      & JSC   & fixed     & 0         & \checkmark& x         & x         & x         & x         & x         & \checkmark& x         \\
10 & bug-227350                      & JSC   & fixed     & 0         & \checkmark& \checkmark& x         & x         & x         & x         & x         & x        \\
11 & bug-244839 & JSC & fixed & 0 & \checkmark& x & x & x & x & x & x & x \\
12 & bug-245066                     & JSC   & fixed     & 0         & \checkmark& x         & x         & \checkmark& x         & x         & x         & x         \\
13 & bug-254331                     & JSC   & fixed     & 0         & \checkmark& x         & x         & x         & x         & x         & \checkmark& x         \\
14 & bug-254574                     & JSC   & fixed     & 0         & \checkmark& x         & x         & \checkmark& x         & \checkmark& \checkmark& \checkmark\\
15 & bug-255582                     & JSC   & fixed     & 0         & \checkmark& x         & \checkmark& x         & x         & x         & x         & x         \\
16 & bug-255715                     & JSC   & confirmed   & 0         & \checkmark& x         & x       & x         & x         & x         & \checkmark& x         \\
17 & bug-273066 & JSC & duplicate & 0 & \checkmark& x & x & x & x & x & x & x \\
18 & bug-273964 & JSC & fixed & 0 & \checkmark& x & x & x & x & x & x & x \\
19 & bug-273979 & JSC & duplicate & 0 & \checkmark& x & x & x & x & x & x & x \\
20 & bug-274052 & JSC & fixed & 0 & \checkmark& x & x & x & x & x & x & x \\
21 & bug-275014 & JSC & duplicate & 0 & \checkmark& x & x & x & x & x & x & x \\
\cmidrule{1-13}
22 & CVE-2019-5866                  & V8    & fixed     & \$500     & \checkmark& x         & x         & x         & x         & x         & x         & x         \\
23 & CVE-2022-1496                  & V8    & fixed     & \$1,000   & \checkmark& x         & x         & x         & x         & x         & x         & x         \\
24 & CVE-2023-6702                  & V8    & fixed     & \$16,000   & \checkmark& x         & x         & x         & x         & x         & x         & x         \\
25 & bug-40243538                    & V8    & fixed     & 0         & \checkmark         & x         & x         & \checkmark& x         & x         & x         & x         \\
26 & bug-42204376                   & V8    & fixed     & 0   & \checkmark& x         & x         & x         & x         & x         & x         & x         \\
27 & bug-42204462                   & V8    & fixed     & 0   & \checkmark& x         & x         & x         & x         & x         & x         & x         \\
28 & bug-334413740                  & V8    & confirmed     & 0   & \checkmark& x         & x         & x         & x         & x         & x         & x         \\
\cmidrule{1-13}
29 & bug-1886940 & SM & duplicate & 0 & \checkmark& x & x & x & x & x & x & x \\
30 & bug-1899113 & SM & duplicate & 0 & \checkmark& x & x & x & x & x & x & x \\
31 & bug-1902139 & SM & duplicate & 0 & \checkmark& x & x & x & x & x & x & x \\
\cmidrule{1-13}
32 & CVE-2019-1428                  & CH    & fixed     & \$15,000  & \checkmark& \checkmark& -         & -         & x         & x         & \checkmark& \checkmark\\
33 & CVE-2020-0710                  & CH    & fixed     & \$15,000  & \checkmark& x         & -         & -         & x         & x         & x         & x         \\
34 & CVE-2020-0712                  & CH    & fixed     & \$15,000  & \checkmark& x         & -         & -         & x         & x         & \checkmark& x         \\
\cmidrule{1-13}
35 & CVE-2021-41751                 & Jerry & fixed     & 0         & \checkmark& x         & -         & x         & -         & -         & x         & x         \\
36 & CVE-2021-41752                 & Jerry & fixed     & 0         & \checkmark& x         & -         & x         & -         & -         & x         & x         \\
37 & CVE-2023-30414                 & Jerry & submitted    & 0         & \checkmark& \checkmark& -         & x         & -         & -         & \checkmark& \checkmark\\
38 & CVE-2023-30410                 & Jerry & submitted    & 0         & \checkmark         & x         & -         & \checkmark& -         & -         & x         & x         \\
39 & CVE-2023-30408                 & Jerry & submitted    & 0         & \checkmark& \checkmark& -         & x         & -         & -         & \checkmark& \checkmark\\
40 & CVE-2023-30406                 & Jerry & confirmed   & 0         & \checkmark& \checkmark& -         & x         & -         & -         & x         & x         \\
41 & CVE-2024-33257 & Jerry & submitted & 0 & \checkmark & x & - & x & - & - & x & x \\
42 & CVE-2024-33258 & Jerry & fixed     & 0 & \checkmark & \checkmark & - & x & - & - & x & x \\
43 & CVE-2024-33259 & Jerry & submitted & 0 & \checkmark & x & - & x & - & - & x & x \\
44 & CVE-2024-33260 & Jerry & submitted & 0 & \checkmark & x & - & x & - & - & x & x \\
45 & CVE-2024-33254 & Jerry & confirmed & 0 & \checkmark & x & - & x & - & - & x & x \\
46 & CVE-2024-33255 & Jerry & submitted & 0 & \checkmark & x & - & x & - & - & x & x \\
47 & bug-5141 & Jerry & fixed & 0 & \checkmark & x & - & x & - & - & x & x \\
\cmidrule{1-13}
48 & CVE-2024-33262 & QJS & confirmed & 0 & \checkmark& x & - & x & - & - & x & x \\
49 & CVE-2024-33263 & QJS & fixed & 0 & \checkmark& x & - & x & - & - & x & x \\
50 & bug-273 & QJS & confirmed & 0 & \checkmark& x & - & x & - & - & x & x \\
51 & bug-283 & QJS & confirmed & 0 & \checkmark& x & - & x & - & - & x & x \\
52 & bug-431 & QJS & fixed & 0 & \checkmark& x & - & x & - & - & x & x \\
53 & bug-432 & QJS & confirmed & 0 & \checkmark& x & - & x & - & - & x & x \\
54 & bug-433 & QJS & confirmed & 0 & \checkmark& x & - & x & - & - & x & x \\
\cmidrule{1-13}
\multicolumn{4}{r|}{Total}                              & \$62,500  & 54        & 9         & 2         & 5         & 0         & 2         & 10        & 6         \\
\bottomrule
\end{tabular}
\vspace{-5pt}
\end{table*}

%% file: coverage2.tex
\begin{table*}[htbp]
\setlength{\tabcolsep}{3.5pt}
\centering
\footnotesize
\caption{Line coverage (10-campaign average) after 7-days fuzzing by \tool{}, baselines, and variants.}
\label{tab:coverage}
\begin{tabular}{lr|cccccccccc|cccccccc}
\toprule
\multirow{3}[3]{*}{Target}  & \multirow{3}[3]{*}{Metric}    & \multicolumn{10}{c|}{Comparison with baselines} & \multicolumn{6}{c}{Ablation study}\\
\cmidrule{3-18}
& & \multicolumn{2}{c}{\tool{}} & \multicolumn{2}{c}{Superion}& \multicolumn{2}{c}{DIE}& \multicolumn{2}{c}{Fuzzilli} & \multicolumn{2}{c|}{FuzzJIT} & \multicolumn{2}{c}{-Seed} & \multicolumn{2}{c}{-Selective} & \multicolumn{2}{c}{-Mutation} \\
\cmidrule{3-18}
& & All & Patch & All & Patch & All & Patch & All & Patch & All & Patch & All & Patch & All & Patch & All & Patch \\
\cmidrule{1-18}
\multirow{4}{*}{JSC}    
& Average          & \textbf{0.64} & \textbf{0.70} & 0.47 & 0.51 & 0.47 & 0.50 & 0.47 & 0.51 & 0.48 & 0.52 & 0.49 & 0.52 & 0.61 & 0.63 & 0.60 & 0.63  \\
& Improvement         & -             & -             & 0.36 & 0.37 & 0.36 & 0.40 & 0.36 & 0.37 & 0.33 & 0.35 & 0.31 & 0.35 & 0.05 & 0.11 & 0.07 & 0.11  \\
& $\hat{A}_{12}$& -             & -             & 0.99 & 0.99 & 0.99 & 0.99 & 0.99 & 0.99 & 0.99 & 0.99 & 0.99 & 0.99 & 0.99 & 0.99 & 0.99 & 0.99  \\
& $p_U$         & -             & -             & \textless0.01& \textless0.01& \textless0.01& \textless0.01&\textless0.01& \textless0.01& \textless0.01& \textless0.01& \textless0.01& \textless0.01& \textless0.01& \textless0.01& \textless0.01& \textless0.01\\
\cmidrule{1-18}
\multirow{4}{*}{V8} 
& Average          & \textbf{0.33} & \textbf{0.41} & 0.20 & 0.26 & 0.19 & 0.26 & 0.19 & 0.26 & 0.21 & 0.27 & 0.20 & 0.26 & 0.24 & 0.30 & 0.24 & 0.33  \\
& Improvement         & -             & -             & 0.65 & 0.58 & 0.74 & 0.58 & 0.74 & 0.58 & 0.57 & 0.52 & 0.65 & 0.58 & 0.38 & 0.37 & 0.38 & 0.24  \\
& $\hat{A}_{12}$& -             & -             & 0.99 & 0.99 & 0.99 & 0.99 & 0.99 & 0.99 & 0.99 & 0.99 & 0.99 & 0.99 & 0.99 & 0.99 & 0.99 & 0.99  \\
& $p_U$         & -             & -             & \textless0.01& \textless0.01& \textless0.01& \textless0.01&\textless0.01& \textless0.01& \textless0.01& \textless0.01& \textless0.01& \textless0.01& \textless0.01& \textless0.01& \textless0.01& \textless0.01\\
\cmidrule{1-18}
\multirow{4}{*}{SM} 
& Average          & \textbf{0.63} & \textbf{0.80} & 0.39 & 0.52 & 0.38 & 0.49 & 0.44 & 0.56 & 0.46 & 0.56 & 0.45 & 0.49 & 0.59 & 0.70 & 0.58 & 0.70  \\
& Improvement         & -             & -             & 0.62 & 0.54 & 0.66 & 0.63 & 0.43 & 0.43 & 0.37 & 0.43 & 0.40 & 0.63 & 0.07 & 0.14 & 0.09 & 0.14  \\
& $\hat{A}_{12}$& -             & -             & 0.99 & 0.99 & 0.99 & 0.99 & 0.99 & 0.99 & 0.99 & 0.99 & 0.99 & 0.99 & 0.99 & 0.99 & 0.99 & 0.99  \\
& $p_U$         & -             & -             & \textless0.01& \textless0.01& \textless0.01& \textless0.01&\textless0.01& \textless0.01& \textless0.01& \textless0.01& \textless0.01& \textless0.01& \textless0.01& \textless0.01& \textless0.01& \textless0.01\\
\cmidrule{1-18}
\multirow{4}{*}{CH} 
& Average          & \textbf{0.66} & \textbf{0.66} & 0.47 & 0.49 & -    & -    & -    & -    & 0.50 & 0.53 & 0.56 & 0.57 & 0.61 & 0.61 & 0.60 & 0.62  \\
& Improvement         & -             & -             & 0.40 & 0.35 & -    & -    & -    & -    & 0.32 & 0.25 & 0.18 & 0.16 & 0.08 & 0.08 & 0.10 & 0.06  \\
& $\hat{A}_{12}$& -             & -             & 0.99 & 0.99 & -    & -    & -    & -    & 0.99 & 0.99 & 0.99 & 0.99 & 0.99 & 0.99 & 0.99 & 0.99  \\
& $p_U$         & -             & -             & \textless0.01& \textless0.01& -    & -   & -    & -    & \textless0.01& \textless0.01& \textless0.01& \textless0.01& \textless0.01& \textless0.01& \textless0.01& \textless0.01\\
\cmidrule{1-18}
\multirow{4}{*}{Jerry} 
& Average          & \textbf{0.86} & \textbf{0.87} & 0.81 & 0.82 & -    & -    & 0.74 & 0.75 & -    & -    & -    & -    & 0.81 & 0.82 & 0.81 & 0.81  \\
& Improvement         & -             & -             & 0.06 & 0.06 & -    & -    & 0.16 & 0.16 & -    & -    & -    & -    & 0.06 & 0.06 & 0.07 & 0.07  \\
& $\hat{A}_{12}$& -             & -             & 0.99 & 0.99 & -    & -    & 0.99 & 0.99 & -    & -    & -    & -    & 0.99 & 0.99 & 0.99 & 0.99  \\
& $p_U$         & -             & -             & \textless0.01& \textless0.01& -    & -    & \textless0.01& \textless0.01& -    & -    & -    & -    & \textless0.01& \textless0.01& \textless0.01& \textless0.01\\

\cmidrule{1-18}
\multirow{4}{*}{QJS} 
& Average          & \textbf{0.60} & \textbf{0.70} & 0.48 & 0.52 & -    & -    & 0.46 & 0.52 & -    & -    & -    & -    & 0.56 & 0.54 & 0.53 & 0.52   \\
& Improvement         & -             & -             & 0.25 & 0.35 & -    & -    & 0.30 & 0.35 & -    & -    & -    & -    & 0.07 & 0.11 & 0.13 & 0.35  \\
& $\hat{A}_{12}$& -             & -             & 0.99 & 0.99 & -    & -    & 0.99 & 0.99 & -    & -    & -    & -    & 0.99 & 0.99 & 0.99 & 0.99  \\
& $p_U$         & -             & -             & \textless0.01& \textless0.01& -    & -    & \textless0.01& \textless0.01& -    & -    & -    & -    & \textless0.01& \textless0.01& \textless0.01& \textless0.01\\

\cmidrule{1-18}
& Average          & \textbf{0.62} & \textbf{0.69} & 0.47 & 0.52 & 0.35 & 0.42 & 0.46 & 0.52 & 0.41 & 0.47 & 0.43 & 0.46 & 0.57 & 0.60 & 0.56 & 0.60  \\
& Improvement         & -             & -             & 0.32 & 0.33 & 0.77 & 0.64 & 0.35 & 0.33 & 0.51 & 0.47 & 0.44 & 0.50 & 0.09 & 0.03 & 0.11 & 0.15  \\
\bottomrule
\end{tabular}
\end{table*}

%% file: related.tex
\section{Related works}

Our main contributions include recognizing security patches and extracting executable PoCs, hot code guidance for fuzzing, and fuzzing for JavaScript engines. 
Here we present a review of three closely related categories of existing works.

\noindent\textbf{Security patches recognition approaches.}
The literature presents a range of approaches that employ learning-based techniques to train security patch recognizers using labeled git commits. 
Notable examples include SPI~\cite{zhou2021spi}, E-SPI~\cite{wu2022enhancing}, VulCurator~\cite{nguyen2022vulcurator}, and GraphSPD~\cite{wang2022graphspd}, which develop their own deep learning models and train them from scratch.
However, it is worth noting that these approaches often underperform when applied to JavaScript engines. 
We provide a rule-based solution specifically tailored to JavaScript engines, offering a simple yet effective approach.
Other works, like PatchDB~\cite{wang2021patchdb}, focus on expanding existing security patch datasets by identifying wild commits that bear similarities to known patches. 
Some researchers explore alternative sources of information, such as GitHub issues~\cite{nguyen2022hermes,tan2021locating} and developer behaviors within the git repository~\cite{farhi2023detecting}, to identify security patch commits.

\noindent\textbf{Directed greybox fuzzing.}
AFLGo~\cite{bohme2017directed} generates inputs to reach specific target locations. 
SelectFuzz~\cite{luo2022selectfuzz} explores relevant paths to improve crash reproduction and detect vulnerabilities. 
Hawkeye~\cite{chen2018hawkeye} uses static analysis to collect information such as call graph, function, and basic block level distances to target sites. 
These works do not assume the availability of PoCs and rely on relatively heavier static analysis to gradually select the test cases that are closer to the target lines and finally make a coverage there.
They cannot scale to large projects like JavaScript engines.

\noindent\textbf{JavaScript engine fuzzing.}
Several fuzzing tools have been developed to utilize historical PoCs. 
LangFuzz~\cite{holler2012langfuzz} mutates the PoC by replacing subtrees of parsed ASTs to uncover new bugs. 
Superion~\cite{wang2019superion} and Nautilus~\cite{aschermann2019nautilus} enhance LangFuzz by implementing coverage feedback to maintain the middle sample between seeds and test cases that result in crashes. 
DIE~\cite{park2020fuzzing}, Montage~\cite{lee2020montage}, and CodeAlchemist~\cite{han2019codealchemist} improve the validity of generated seeds by preserving the original structure and variable types during mutation, using an LSTM model, and utilizing constraints of variable types respectively. 
ComFuzz~\cite{ye2023generative} fine-tunes a pre-trained model for completing JavaScript code snippets using a given prefix.
It creates new tests based on the headers of historical ones, then applies additional mutations to enhance their ability to reveal bugs, leading to more substantial alterations to the original PoCs.
Many of these methods have overly focused on the strategy of mutation and neglected the significance of seeds. 
Outdated seed collections are utilized by a few~\cite{wang2019superion}, while others directly employ unprocessed PoCs as seeds~\cite{lee2020montage,han2019codealchemist}. 
Some even fragment the seeds and/or reassemble them entirely~\cite{lee2020montage, ye2023generative}. 
In addition to acquiring and processing premium quality seeds, our approach is customized to optimize the effectiveness of historical PoCs.
In fact, the PoCs collected by \tool{} can be leveraged by any fuzzing tools relying on seeds to further boost their performance.
Recently, some other powerful JavaScript engine fuzzers, relying on no seeds, have been devised, including  Fuzzilli~\cite{gross2018fuzzil} and FuzzJIT~\cite{wangfuzzjit}.
Fuzzilli proposes more fine-grained mutation on the intermediate language than AST.
FuzzJIT detects optimization bugs by comparing whether the optimization changed the semantics of optimized code.

%% file: conclude.tex
\section{Conclusion and future work}

In this work, we propose a novel JavaScript engine fuzzing approach based on security patches, named \tool{}. 
\tool{} first recognizes security patches in the given JavaScript engine and extracts PoCs from the recognized security patches as seeds. 
The seeds are mutated to generate a large number of new samples. 
At the same time, \tool{} extracts a list of hot code functions from the security patches and selectively inserts probes into the test target. 
We use the instrumented test targets to execute the generated new samples to see if they trigger new crashes or code coverage. 
\tool{} is the first patch fuzzer that leverages both the PoCs and code revisions from security patches.
It successfully found 54 vulnerabilities in six mainstream JavaScript engines, among which 25 obtained CVE IDs and a total of \$62,500 bounty is received.
For future research directions, we aim to explore the applicability of our approach to general open-source projects, extending beyond JavaScript engines. 

%% file: data.tex
\section{Data Availability}
\noindent \tool{} is open sourced at \url{https://github.com/PatchFuzz/patchfuzz}.